# Does the arXiv lead to higher citations and reduced publisher downloads for mathematics articles?

**[final manuscript, May 2, 2006]**

**Please cite:** *Scientometrics* Vol. 71, No. 2. (May, 2007)

Philip M. Davis
Life Sciences Librarian
Cornell University Library
Ithaca NY, USA
pmd8@cornell.edu

Michael J. Fromerth
arXiv Administrator
Cornell University Library
Ithaca NY, USA
mjf48@cornell.edu

## Abstract

An analysis of 2,765 articles published in four math journals from 1997 to 2005 indicate that articles deposited in the arXiv received 35% more citations on average than non-deposited articles (an advantage of about 1.1 citations per article), and that this difference was most pronounced for highly-cited articles.  Open Access, Early View, and Quality Differential were examined as three non-exclusive postulates for explaining the citation advantage.  There was little support for a universal Open Access explanation, and no empirical support for Early View.  There was some inferential support for a Quality Differential brought about by more highly-citable articles being deposited in the arXiv.  In spite of their citation advantage, arXiv-deposited articles received 23% fewer downloads from the publisher's website (about 10 fewer downloads per article) in all but the most recent two years after publication.  The data suggest that arXiv and the publisher's website may be fulfilling distinct functional needs of the reader.

# Introduction

The arXiv (arXiv.org) provides free and open access to over a third of a million e-prints in physics, mathematics, computer science and quantitative biology. In production since 1991, the arXiv provides a mechanism for researchers to disseminate manuscripts and published articles to a wide community of researchers. Adopted extensively by some of the disciplines it serves, the arXiv is believed to enhance the scholarly communication process by improving access to research findings – often in advance of formal publication – and by providing a platform for the open peer-review of manuscripts, many of which are submitted for formal review and publication in established journals.

It is widely believed that depositing one's manuscripts and articles into an institutional or subject-based archive increases the number of citations to one's article. In computer science, Steve Lawrence illustrated that freely available online articles were cited more often than 'offline' articles published in the same venue [1]. In an analysis of articles published in *The Astrophysical Journal* in 1999 and 2002, Schwarz and Kennicutt reported that papers posted on astro-ph (a section of the arXiv) were cited more than twice as often as those that were not [2]. Travis Metcalfe was able to generalize these results to thirteen different journals, whose authors posted papers on astro-ph [3]. He also reported that astronomy papers published in high-impact science journals such as *Science* and *Nature* received an even larger citation advantage when their authors posted their papers on astro-ph. Kristin Antelman confirmed a citation advantage for top journals in philosophy, political science, electric and electronic engineering and mathematics, when fulltext versions (including drafts, preprints and postprints) were available anywhere on the Internet [4].

Michael Kurtz and others, working with data from the NASA Astrophysics Data System, attempted to determine the cause(s) of the citation advantage, investigating three non-exclusive explanations [5]:

1. **Open Access Postulate.** The arXiv, like other open article repositories, increases access to articles, especially for readers who do not have access to the publisher's subscription-based content.

2) **Early View Postulate.** Articles appear earlier in the arXiv than on the publisher's website, giving author manuscripts a longer period in which to be read and cited.

3) **Self-Selection Postulate.** Authors selectively deposit their best articles in the arXiv. Alternatively, better authors may be using the arXiv (Author Postulate). Both of these explanations, which describe two different behavioral phenomena, essentially describe a quality differential (**Quality Postulate**). The Quality Postulate will be tested in this study.

Kurtz's analysis of astrophysics journals provided support for the Early Access and Self-Selection explanations, but not for Open Access. The authors argued that Open Access as an explanatory cause for increased citations may be overstated, especially for well-funded disciplines, such as astronomy, where other factors like access to equipment and funds are stronger barriers to authors than access to the literature.

While the majority of scientific publishers currently allow authors to post publicly some version of their manuscript – either on personal, institutional, or subject-based websites such as the arXiv – some fear that this will lead to decreased article readings from the publisher's website.

This study sets out to answer two important questions: 1) Do articles deposited in the arXiv receive more citations than non-deposited articles, and 2) Are articles deposited in the arXiv associated with fewer downloads from the publisher's website? In addition to analyzing these associations, the authors will attempt to discern the likely causes of any detectable differences.

## Data and Methods

This study is based on article-level data provided by the London Mathematical Society (LMS) for three of their journals, the *Proceedings*, *Journal*, and *Bulletin* of the LMS, and for the journal, *Compositio Mathematica* published by the LMS on behalf of the Foundation Compositio Mathematica. All four of these journals are published electronically on the Cambridge University Press (CUP) website. The dataset contains research articles (editorials, obituaries, book reviews and similar items were excluded) published between January 1997 and September 2005.

The five variables analyzed in this study were the number of MathSciNet citations, the number of publisher fulltext downloads, the number of arXiv fulltext downloads, the number of publisher viewing days (the number of days that an article has been available online from the publisher's website), and the number of prepublication days (the number of days that an article deposited in the arXiv predates the formal publication). MathSciNet citations were used instead of ISI citations since journal coverage in the latter is incomplete, and because MathSciNet tracks citations to preprints in the arXiv in addition to citations to the formal publication.

While articles published back to 1997 are available from the publisher's website, online usage statistics for LMS journals have been made available only within the last three years. In addition, *Compositio Mathematica* moved from the Kluwer website to the CUP website in 2004, providing less than a year and a half of publisher usage data. These differences, and their impact on the results of this study, will be addressed in the Discussion section. It is not known whether the articles published in these four journals have been deposited in any other subject-based, personal, or institutional archive.

Descriptive statistics were employed in this study, along with linear regression, correlation, and the Mann-Whitney U test – a non-parametric version of the t-test, that compares the rank sums of two samples to determine if they come from the same distribution [6, 7]. Linear regression was used to discern whether certain variables have explanatory power, after controlling for other possible cofactors. For example, does the number of prepublication days affect the number of citations an article receives, controlling for the number of days since it was published and the journal in which it was published?

Because linear regression assumes a normal distribution and equal variance, certain variables required transformation. For citations, the transformation was log(citations +1), because over a third of the articles were not cited at the time of this study, and the log of zero is a logical impossibility. For downloads, a simple log transformation was required. In both cases, these transformations were sufficient for achieving normality and equal variance.

## Results

From a total of 2,765 articles published in the four journals between 1997 and 2005, 511 (18.5%) were found in the arXiv, whereas 2,254 (81.5%) were not. The inclusion rate ranged from 11.2% for the *Journal* to 34.3% for *Composito Mathematica* (Table 1). The number of LMS articles deposited in the arXiv has been gradually rising, from only 7.6% in 1997 to nearly 37% in 2005 (Table 2). Since 1997, the arXiv adoption rate has grown for all four journals. In 2005, the adoption rate was 35% (*Proceedings*), 23% (*Journal*), 26% (*Bulletin*), and 64% (*Composito Mathematica*).

**Citations**

Articles deposited in the arXiv received significantly more citations than non-deposited articles. In each year since 1997, the mean number of citations to articles deposited in the arXiv was greater than the mean number of citations to articles not found in the arXiv (Table 2 and Figure 1). This difference was as large as 2.1 citations per article in 1997 and as low as 0.8 citations per article in 2001. The mean difference over all years was 1.1 citations per article (or a 35% difference) in favor of articles deposited in the arXiv.

Table 3 provides a comparative breakdown of citations. Nearly a third of the articles received no citations (32.3% of articles deposited in the arXiv compared to 32.9%). At the same time, articles with more than 5 citations were more likely to be found within the arXiv group than the non-arXiv group (14.9% versus 12.3% respectively).

These differences are more profound when one realizes that articles found in the arXiv have been published more recently. The mean number of published days for articles deposited in the arXiv was 1,142 days (approx. 3.1 years), compared to 1,666 days (approx. 4.6 years). For articles that received more than 5 citations, the mean number of published days was 1,780 (approx. 4.9 years) for arXiv deposited articles, compared to 2,205 days (approx 6.0 years) for non-deposited articles. In sum, the citation differential appears to be greater for highly-cited articles in the arXiv despite the fact that they were published more recently.

**Fulltext Downloads**

Articles deposited in the arXiv also received significantly fewer fulltext downloads from the publisher's website in virtually all years except for 2004 and 2005 (Table 2). Articles in the arXiv received about 23% fewer article downloads on average (about 10 downloads per article), and this difference was much higher for older articles than for newer ones. Articles published in 2004 and 2005 received statistically identical fulltext downloads irrespective of whether the article was deposited in the arXiv. The reader is reminded that the publisher began counting article downloads in 2003, so article downloads for older articles do not include initial readership when the article was first published (see Figure 2).

## Discussion

The results of this study show that in general, articles deposited in the arXiv receive more citations and fewer downloads at the publishers website than non-deposited articles. Since it is impossible to do a controlled scientific study, explanatory causes must be inferred by systematically ruling out all other plausible causes. The three non-exclusive postulates, Open Access, Early View, and Quality Differential are analyzed below.

**Open Access Postulate**

If increased access were responsible, at least in part, for a citation advantage, one would expect to find a significant and positive relationship between fulltext arXiv downloads and the number of citations an article receives – the theory being that additional readership from the arXiv is responsible for additional citations.

Although a significant and positive correlation between arXiv downloads and article citations does exist for the entire dataset, the relationship is rather weak (R=0.15). A LOESS curve was fitted to the data (Figure 3) so that the relationship between these two variables could be better understood. LOESS (locally weighted regression) uses the same least squares method as regression but on localized subsets of the data. LOESS fits a curve to the data without any preconceived notions of its shape, and is thus an excellent procedure to explore complex relationships [8, 9].

According to Figure 3, there appears to be no relationship between fulltext arXiv downloads and the number of article citations until an article has been downloaded more than 400 times (log downloads > 2.6). These results are consistent with the detailed work of Henk Moed on the journal *Tetrahedron Letters*, which suggests that a relatively small group of both highly cited and frequently downloaded papers are responsible for the weak, though statistically significant, correlation between downloads and citations [10].

In summary, if the Open Access postulate is a valid explanation for the citation advantage, its effect may be severely limited to highly-cited articles.

**Early View Postulate**

If the Early View explanation is at least partially responsible for our results, one would expect that articles deposited in the arXiv farther in advance of their formal publication date would receive more citations, since the longer an article is available for public viewing, the greater probability that it will be read and cited. The Early View postulate can be tested by regressing the number of prepublication days for articles in the arXiv on their log number of MathSci Net citations while controlling for both the number of days since publication and the journal. Such an analysis indicates that the number of prepublication days does not provide any significant explanatory power in accounting for the citation advantage ($B_{prepublication}$ = 1.73 x $10^{-5}$, P=0.59). A histogram of prepublication days is presented as Figure 4, and a partial regression plot is presented as Figure 5. In addition, those articles that were deposited *after* the publisher version was released (indicted in Figure 4 by the ^ ) received more citations than those deposited as prepublications (4.2 versus 2.6, P<0.05). These findings provide support for explanations other than the Early View Postulate.

**Quality Differential Postulate**

In "Online or Invisible", the report that became the seminal letter published in *Nature*, Steve Lawrence wrote about the difficulty of ascribing causation on the basis of his findings. He states, "Online articles may be more highly cited because they are easier to access and thus more visible and more likely to be read, or because higher quality articles are more likely to be made available online", implying that both an Open Access and a Quality Differential postulate are valid [11]. Comparing articles published in the same journal, Lawrence argued that Open Access may be the stronger explanation for computer conference papers. The approach of comparing articles published in the same journal was also used by Stevan Harnad and Tim Brody in their work on physics articles [12]. While this approach attempts to control for the quality difference between journals, it assumes equal quality across articles published in the same journal. In fact, however, the distributions of citations is highly skewed within each journal; approximately 15% of the articles in a journal collect 50% of the citations, and half of the articles collect about 90% [13]. In our dataset, 12% of the published articles received 50% of the total citations, and 44% of the articles received 90%.

If authors are indeed depositing their best papers in the arXiv, Open Access may be merely an artifact, not the cause, of a citation differential. Table 3 shows that a higher percentage of highly-cited articles (receiving more than 5 citations) are found in the arXiv, in spite of being significantly younger than non-deposited articles. In addition, we have failed to find support for the Early View postulate and have provided evidence that many highly-cited articles were deposited into the arXiv *after* formal publication. Both of these findings lend support to the Quality Differential postulate.

Michael Kurtz's analysis of astrophysics articles deposited in the arXiv reported strong evidence for the Quality and Early View postulates but not Open Access [5]. Jonathan Wren's analysis of biomedical journals illustrated that articles from higher impact journals were more frequently found at non-journal websites. As a behavioral explanation, he described a "trophy effect", essentially a desire for researchers to self-promote and display their own accomplishments [14]. Wren's explanation also lends support for the Quality Differential postulate.

In sum, there is very little support for a universal Open Access explanation. If Open Access is responsible for the citation differential between arXiv-deposited and non-deposited articles, it is severely limited to highly-cited articles. There was no empirical evidence for the Early View postulate and inferential support for some form of a Quality Differential. It is not known whether articles deposited in the arXiv were also deposited in other disciplinary or institutional archives, or are found on the author's website. A more complete investigation would help better understand how the arXiv is used by researchers to promote their work.

**Explanation for the reduction in downloads from the publisher's site**

As Figure 2 illustrates, articles deposited in the arXiv are less often downloaded from the publisher's website. The most plausible explanation is that the arXiv copy is sufficient, in many cases, for the purposes of the reader. Some of the readers who turn to the arXiv instead of the publisher's website may not have online access to these math journals. However, the documented phenomenon that needs to be explained is not the rise in arXiv downloads, but the *decline in publisher downloads*. The fact that arXiv-deposited articles are cited *more* than other articles published in the same journals rules out the possibility that arXiv-deposited articles are of lower quality and thus are viewed less often from the publisher's website. If citations are a measure of quality, one would expect *greater* (not fewer) downloads from the publisher's website. Clearly, the most plausible explanation for the decrease is the presence a nearly-identical article in the arXiv. The policy of the London Mathematical Society is to permit any version of a preprint (but not a proof) to be posted publicly. The LMS also highly recommends that manuscripts are submitted in TeX, a document preparation system for typesetting scientific and technical manuscripts which has features of formatting mathematical formulae. For the purposes of the mathematician, a final peer-reviewed preprint including correctly formatted formulae may be nearly as good as a final published copy.

Table 2 and Figure 2 do not suggest that readers are wholesale turning away from the publisher's final version in lieu of an arXiv copy, however. Cumulative publisher downloads for articles published in 2004 and 2005 were statistically identical for deposited and non-deposited articles, and this observation was not consistent with the large decreases experienced for all other years. The most likely explanation is that the arXiv and publisher's website fulfill different functional needs. The publisher's website may be better for information discovery and browsing, especially for recently published articles. In contrast, the arXiv may provide some competition for known article searches. ArXiv articles have short, permanent URL (i.e. http://arxiv.org/abs/cs.DL/0603056), which may make them faster to retrieve and easier to share with one's colleagues.

Henk Moed's detailed study of downloads and citations describe two usage factors in describing article downloads: an "ephemeral factor", which represents the initial browsing during the first three months of an article's publication, and a "residual factor", which represents the focused use of older (or "archival") articles [10]. These processes are not mutually exclusive, but overlapping, and can describe the behavior of a single individual in the same session. Qualitative studies of readers and their rationale for choosing either the arXiv or the publisher's website may help answer if and how these two types of resources fulfill different functional needs of the reader.

### **Generalizability and further study**
This study focused on four journals published by a single mathematics society. Because of the limited sample size, care should be taken not to generalize very broadly. The results, however, may be representative of other journals in mathematics, and further studies are needed to confirm these results.

In many ways, mathematics journals are more like those in the humanities literature than those in the sciences. Most math journals have a citation half-life of more than ten years, and while this study covered a period of nine years, the majority of articles that were submitted to the arXiv were placed there over the last five years. While other studies in physics and astronomy have detected a larger citation differential, the nature of mathematics may require a longer period of observation.

Further studies should focus on other disciplines, like physics, that have adopted the arXiv as a standard tool for the communication of research results. A larger dataset composed of more journals would be able to more accurately investigate how the prestige of a journal affects the citations of its article. Lastly, a multiple-publisher study would be able to control the effect of a publisher's interface on full-text article views. Recent work by the author suggests that a publisher's interface can have dramatic effect on article usage patterns even when controlling for the same journal content [15].

### **Acknowledgements**
The authors wish to thank Susan Hezlet of the London Mathematical Society for providing the data used in this study. Sincere gratitude also goes to Paul Ginsparg, Simeon Warner, Bill Walters, Karla Hahn, David Goodman and Steve Rockey for their feedback. This study would not be possible without the administrative support of the Cornell University Library. The positions held in this paper are entirely those of the authors and do not necessarily reflect those of the LMS or the Cornell University Library.

# Tables and Figures

**Table 1. Summary statistics by journal**

| Journal | Articles Published | Number of articles in arXiv |
|---|---|---|
| Proceedings of the LMS | 464 | 95 (20.5%) |
| Journal of the LMS | 955 | 107 (11.2%) |
| Bulletin of the LMS | 713 | 92 (12.9%) |
| Composito Mathematica | 633 | 217 (34.3%) |
| Total | 2765 | 511 (18.5%) |

**Table 2. Differences in citations and downloads by year**

| Year | Articles Published | Number of articles in arXiv | Mean difference in number of article citations (arXiv - non) | Mean difference in number of fulltext downloads from publisher's website (arXiv - non) |
|---|---|---|---|---|
| 1997 | 301 | 23 (7.6%) | 2.1 (37%) * | -15.9 (-37%) * |
| 1998 | 282 | 26 (9.2%) | 1.6 (31%)* | -17.3 (-46%) ** |
| 1999 | 341 | 43 (12.6%) | 1.3 (28%) | -4.4 (-12%) |
| 2000 | 345 | 41 (11.9%) | 1.4 (34%) * | -10.5 (-26%) ** |
| 2001 | 315 | 51 (16.2%) | 0.8 (21%) * | -10.8 (-27%) ** |
| 2002 | 299 | 63 (21.1%) | 1.2 (40%) * | -20 (-40%) *** |
| 2003 | 324 | 73 (22.5%) | 1.2 (47%) *** | -12.3 (-19%) ** |
| 2004 | 324 | 105 (32.4%) | 0.6 (43%) * | -3.0 (-4%) |
| 2005 | 234‡ | 86 (36.8%) | 0.1 (35%) | 0.9 (2%) |
| Total | 2765 | 511 (18.5%) | mean dif.=1.1 (35%) | mean dif.= -10.4 (-23%) |

‡ Articles published through Sept. 2005
* Significant at P<0.05 (2-tailed Mann-Whitney test)
** Significant at P<0.01 (2-tailed Mann-Whitney test)
*** Significant at P<0.001 (2-tailed Mann-Whitney test)

**Table 3. Distribution of MathSciNet citations**

|  | Article in arXiv? | | |
|---|---|---|---|
|  | no | yes | total |
| Citations | No. (%) | No. (%) | No. (%) |
| 0 | 741 (32.9%) | 165 (32.3%) | 906 (32.8%) |
| 1 | 501 (22.2%) | 104 (20.4%) | 605 (21.9%) |
| 2 | 297 (13.2%) | 58 (11.4%) | 355 (12.8%) |
| 3 | 208 (9.2%) | 62 (12.1%) | 270 (9.8%) |
| 4 | 133 (5.9%) | 31 (6.1%) | 164 (5.9%) |
| 5 | 96 (4.3%) | 15 (2.9%) | 111 (4.0%) |
| >5 | 278 (12.3%) | 76 (14.9%) | 354 (12.8%) |
| **Total** | 2254 (100%) | 511 (100%) | 2765 (100%) |

**Figure 1. Mean MathSciNet citations (± 95% Confidence Interval) for each year**

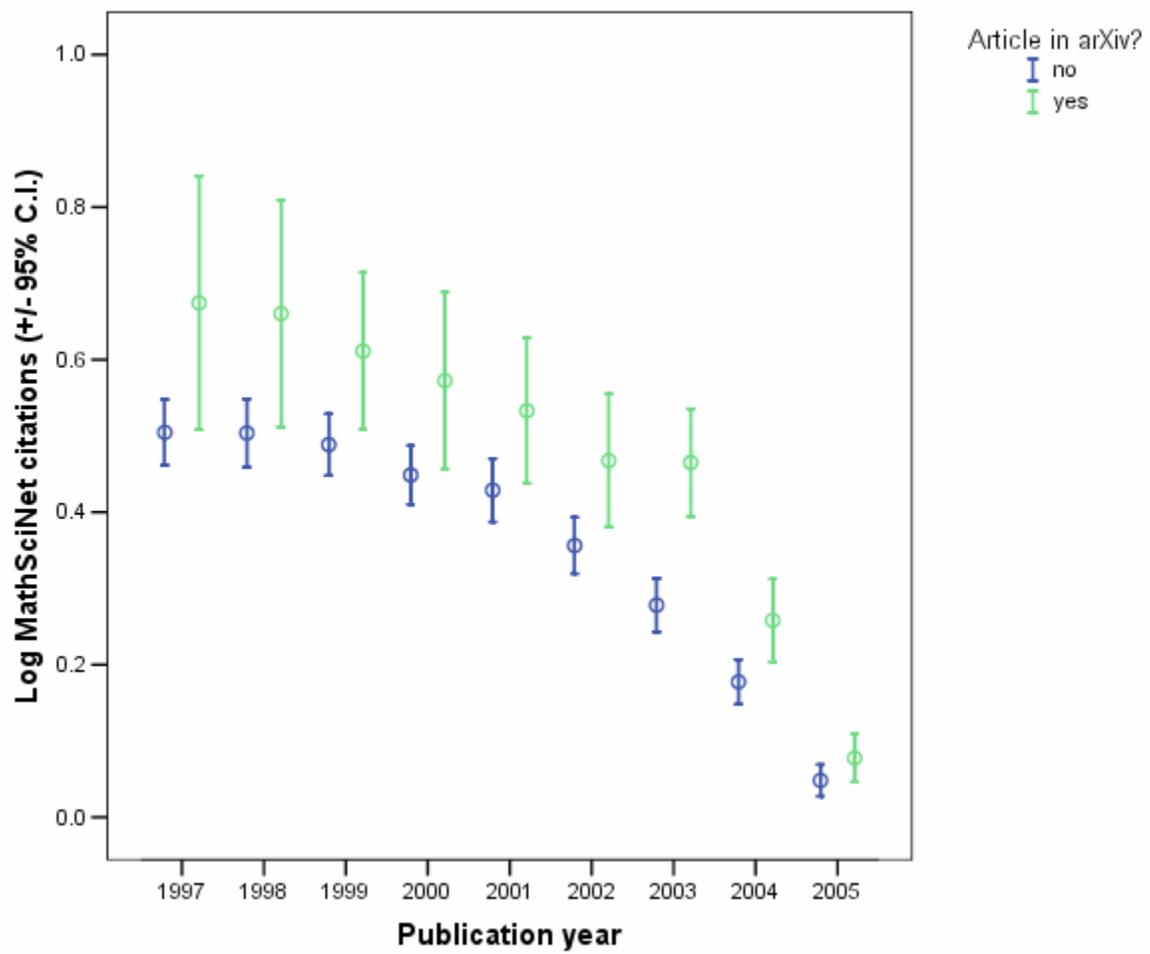

**Figure 2. Mean publisher fulltext downloads (± 95% Confidence Interval) for each year**

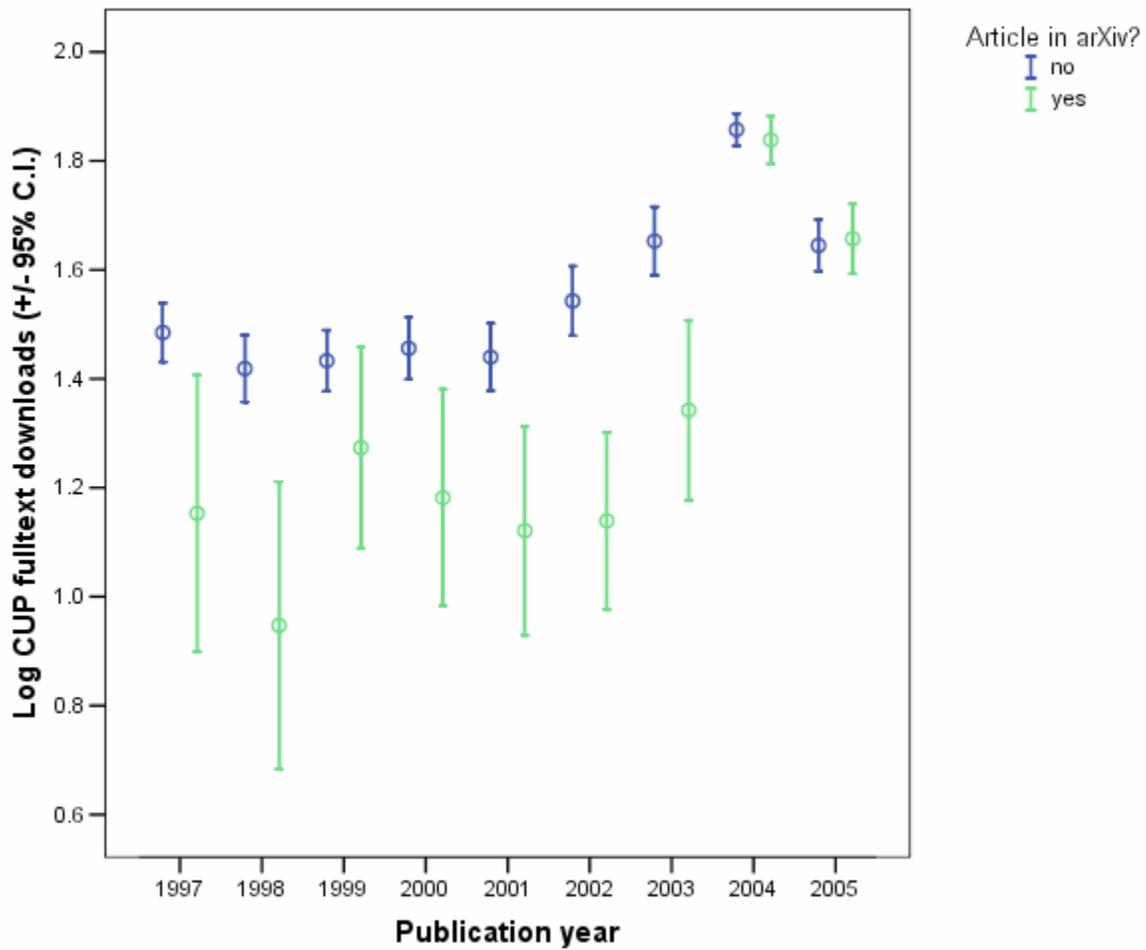

Note: The publisher began counting fulltext downloads in 2003. The use of these older articles does not therefore include initial readings when the article was published, and explains the shape of this graph.

**Figure 3. Relationship between arXiv downloads and citations with fitted LOESS curve (α=0.5)**

**Figure 4. Distribution of arXiv submissions by number of days before formal publication (i.e. prepublication days)**

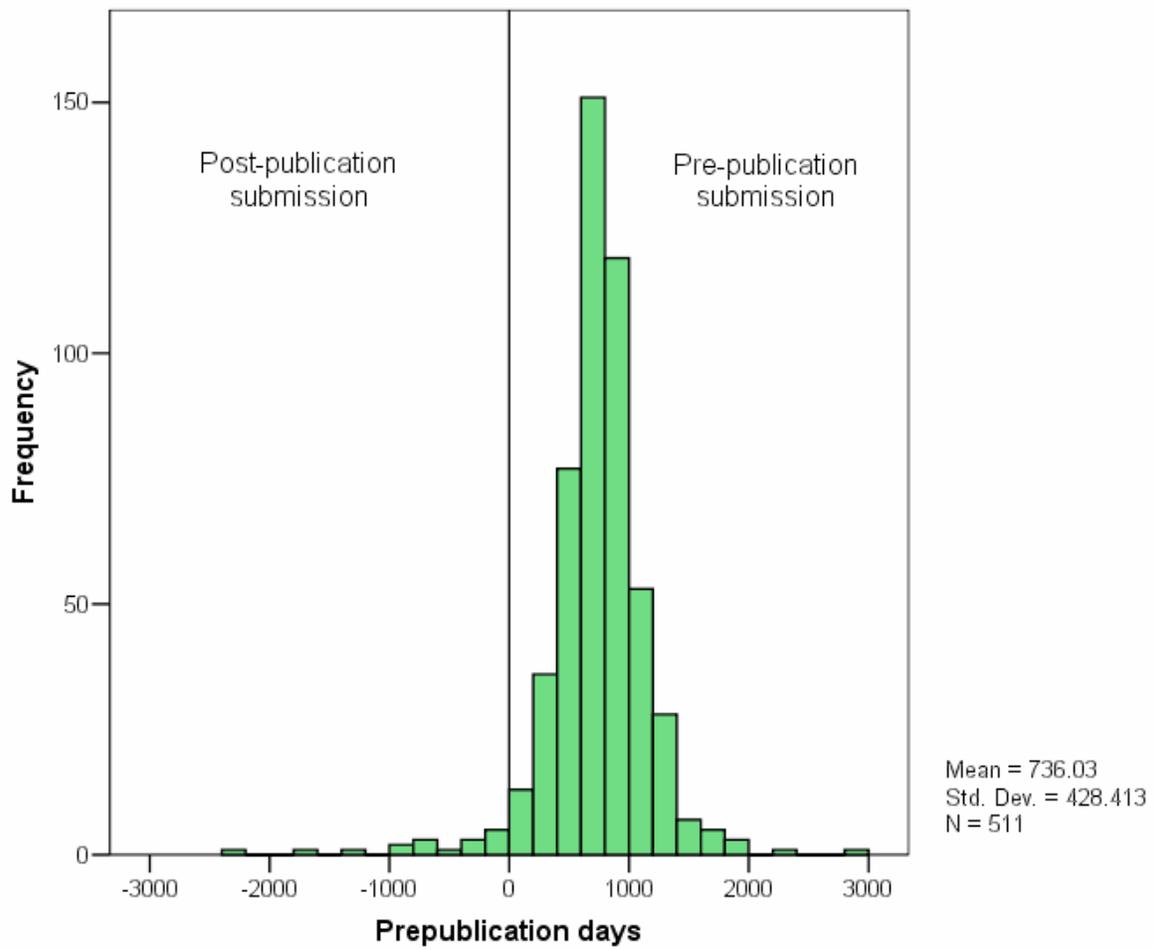

**Figure 5. Partial regression plot of prepublication days on citation, controlling for the number of days since publication and the effect of the journal. Articles deposited in the arXiv after formal publication are highlighted by ^.**

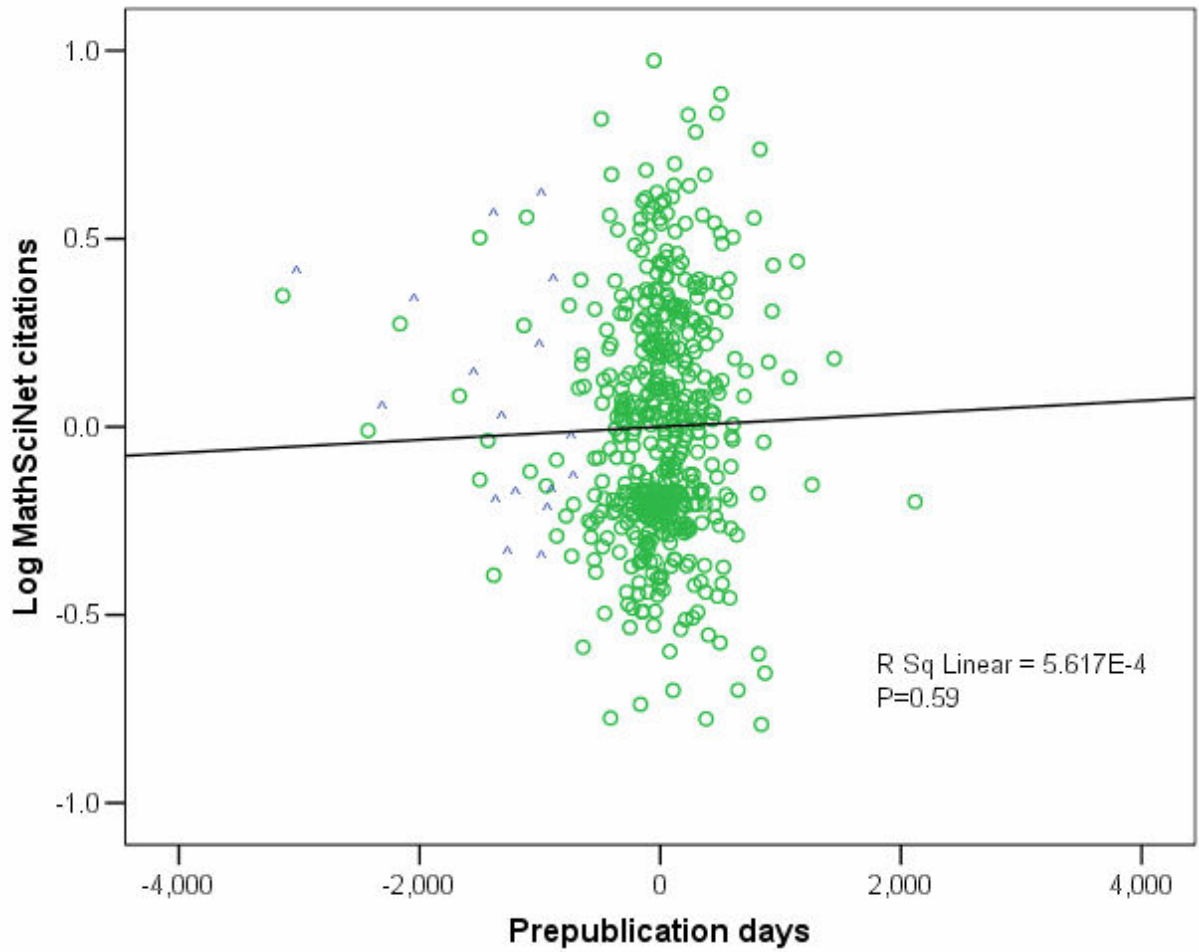